\documentclass[twocolumn,aps,amsmath,amssymb]{revtex4}
\usepackage{graphicx}
\usepackage{dcolumn}
\usepackage{bm}
\begin{document}
\title{Octahedral Tilt Instability of ReO$_3$-type Crystals}
\author{Philip B. Allen}
\affiliation{Department of Physics and Astronomy,
State University of New York, Stony Brook, NY 11794-3800
(permanent address)\\
and
Department of Applied Physics and Applied Mathematics
Columbia University, New York, NY 10032}
\author{Yiing-Rei Chen}
\altaffiliation{current address: Dept. of Chemistry, 
Columbia University, New York, NY 10032}
\affiliation{Department of Physics and Astronomy
State University of New York, Stony Brook, NY 11794-3800}
\author{Santanu Chaudhuri}
\altaffiliation{current address: Dept. of Chemistry, 
Brookhaven National Laboratory, Upton, NY 11973}
\author{Clare P. Grey}
\affiliation{Department of Chemistry
State University of New York, Stony Brook, NY 11794-3400}
\date{\today}

\begin{abstract}
The octahedron tilt transitions of
ABX$_3$ perovskite-structure materials
lead to an anti-polar arrangement of dipoles,
with the low temperature ($T$) structure
having six sublattices polarized along
various crystallographic directions.  It
is shown that an important mechanism driving
the transition is long range
dipole-dipole forces acting on both displacive
and induced parts of the anion dipole.  
This acts in concert with
short range repulsion, 
allowing a gain of 
electrostatic (Madelung) energy, 
both dipole-dipole and charge-charge, because the
unit cell shrinks when the hard ionic spheres
of the rigid octahedron tilt out of linear
alignment.
\end{abstract}

\maketitle


In 1950 Slater \cite{Slater} presented an electrostatic
theory of the ferroelectric transition in BaTiO$_3$
(perovskite ABX$_3$ structure)
by generalizing the Clausius-Mossotti (CM)
picture.  An important ingredient is the fact that the 
local dipolar electric field
$\vec{F}_D$ at the X site differs from the Lorentz
value $(4\pi/3)\vec{P}$ because the local symmetry is
less than cubic.  Here we present a similar discussion,
also highlighting the role of dipole-dipole interactions,
of the octahedron tilt commonly found in perovskite
structure.  To simplify the many interactions,
this paper is confined to the ReO$_3$ structure
type \cite{Daniel} with the perovskite A sublattice empty.  
We take AlF$_3$ as our prototype.

\par
\begin{figure}[t]
\centerline{\scalebox{0.4}[0.4]{\includegraphics{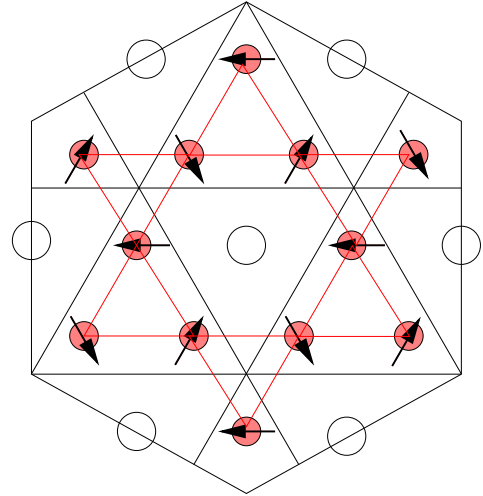}}}
\caption{\label{fig:111}
The (111) planes of perovskite are alternately AX$_3$ and B layers.
The AX$_3$ sites constitute an {\it fcc} lattice with
close-packed triangular (111) planes occupied 75\% by
X anions (shown as filled circles) and 25\% by A cations (shown as
open circles and missing in the AlF$_3$ structure.)
Arrows denote displaced X anions.
Faint triangles with counter-clockwise rotations have their
nearest B cation in the plane below, and inverted faint triangles
with clockwise rotations have their nearest B cation in the plane above.}
\end{figure}
\par

Starting with the parent cubic structure, the
($R\bar{3}c$ rhombohedral) low $T$ phase of AlF$_3$
is generated by a rigid rotation of an AlF$_6$ 
octahedron through angle $\omega\approx 0.3$rad
around a cubic (111) axis through an Al atom.
Because neighboring octahedra share corners, 
rotations alternate, doubling the unit cell
according to wavevector $(\pi,\pi,\pi)$.  
A schematic view of the distorted (111) plane
is shown in Fig. \ref{fig:111}.
The Shannon ionic radii \cite{Shannon} of Al$^{3+}$
and F$^-$ match almost perfectly to an octahedron
with the central Al touching the six F anions, and
each F anion touching its eight F neighbors.
Thus the rigidity of the octahedron
follows both from Al-F covalency and from ionic 
size effects.  As the octahedra tilt, their spacing
shrinks, generating a rhombic primitive cell whose
$c/a$ ratio increases by the factor $1/\cos(\omega)$
relative to the cubic value, and cell volume $V$
decreases as $\Delta V/V=-\sin^2\omega$.

\par
\begin{figure}[t]
\centerline{\scalebox{0.4}[0.4]{\includegraphics{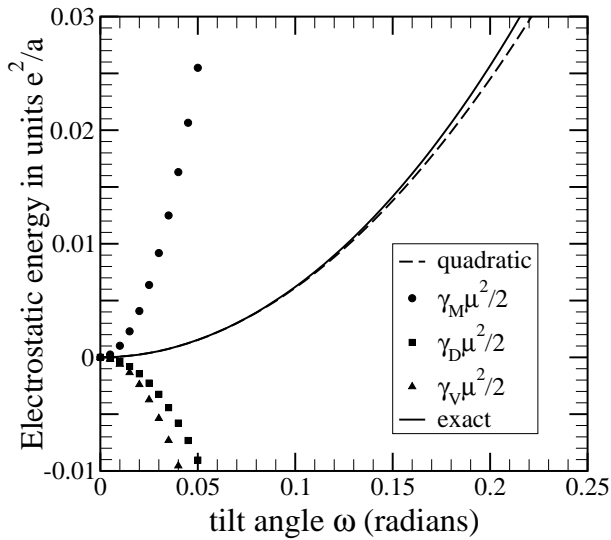}}}
\caption{\label{fig:mad}
Electrostatic energy of BX$_3$ with full ionic charges +3
and -1, as a function of the tilt angle $\omega$ around the (111) axis.
The dashed curve is the full quadratic approximation, namely
the sum of the Madelung ($M$), dipole-dipole ($D$), and
volume ($V$) terms.  
The solid curve is the exact Madelung sum.}
\end{figure}

CM theory \cite{Jackson,Jona} shows that a cubic lattice
of polarizable molecules may have a ``polarization catastrophe''
signalling an instability towards ferroelectric
polarization.  The condition for instability is an
increase of the product $n\alpha$ (density times polarizability)
to $3/4\pi$.  A generalized version of
this statement is derived in the Appendix:
a self-stabilized spontaneously electrically
polarized state will occur when $\alpha$ increases
to $1/\gamma_{\rm max}$ where $\gamma_{\rm max}$ is
the maximum eigenvalue of the dipole-dipole interaction
tensor $\Gamma$.
The tensor $\Gamma_{i\alpha,j\beta}$ is defined as the $\alpha$
Cartesian component of the dipolar electric field $\vec{F}_{D,i}$
(at site $\vec{R}_i$ of the lattice) created by a unit
dipole in the Cartesian $\beta$ direction at site $\vec{R}_j$.
In a $3-N$-vector notation, $|F_D>=\Gamma|\mu>$.
This generalization of Clausius-Mossotti theory requires
no restrictions on the size or
symmetry of the system.  The pattern of
spontaneous polarization, given by the corresponding eigenvector,
is in general not a simple ferroelectric.

Rigid rotations of octahedra almost always occur in the low-$T$
phases of perovskites \cite{Glazer,Woodward}.
AlF$_3$ has the perovskite structure with the A sublattice empty.
Below 730K the structure is rhombohedral because of 
cooperative rotation of AlF$_6$ octahedra, as
shown in Fig. \ref{fig:111}.  
We have successfully modelled this instability in two different ways:
(1) using density-functional theory (DFT) \cite{Chen}, and (2) using classical
molecular dynamics (MD) \cite{Chaudhuri}, including both short-range and 
electrostatic forces, plus anion polarizability.
In this paper we abstract from our earlier results a simple
picture that includes electrostatic effects of charged and polarizable
point ions.

The dipole patterns $|\mu>$ ($<i\alpha|\mu>$ is the $\alpha$ Cartesian
component of the dipole $\vec{\mu}_i$ on the $i$'th ion)
are particularly simple and illuminating.
First define the ``displacive dipole'' $\vec{\mu}_{D,i}$ of the $i$th X anion
as $\vec{\mu}_{D,i}=-\beta e \vec{u}_i$
where $\vec{u}_i$ is the small displacement from the site $\vec{R}_i$ of
cubic symmetry, and $-\beta e$ is the charge of the anion.
The unknown dimensionless parameter $\beta$ absorbs the uncertainty about
what actual charge to assign.  The dipole can be imagined
as a separated pair of opposite charges, the negative end at
the anion nuclear coordinate $\vec{R}_i + \vec{u}_i$, and the
positive end at the lattice site $\vec{R}_i$ formerly occupied by the anion.

Consider the dipole patterns
generated by different types of octahedron tilts.  In
Glazer's \cite{Glazer} notation, when the tilt angle is small,
any tilt $\phi^p \psi^q \theta^r$ is a product of tilts around $\hat{x}$,
$\hat{y}$, and $\hat{z}$ where $p,q,r$ are + or --.  There are
six primitive rotations.  For example,
a $\hat{z}$-tilt by angle $\theta$ is denoted as $\phi^0 \psi^0 \theta^{\pm}$,
where $p=q=0$ indicates no rotation around $\hat{x}$ or $\hat{y}$.
This tilt belongs to wavevector $\vec{k} = (\pi,\pi,k_z)$.
Rotations of adjacent octahedra in the $\hat{x}$ and $\hat{y}$ directions
are forced to be opposite in sign (wavevector $k_x=k_y=\pi$), whereas
in the $\hat{z}$-direction, the next octahedron can have a different
rotation angle.  Glaser's conventions are that $r=+$ corresponds
to $k_z=0$, while $r=-$ indicates $k_z=\pi$.
We have discovered the interesting fact, explained in the Appendix,
that the corresponding dipole patterns $|\mu>$ for the six primitive
rotations are eigenvectors of $\Gamma$.  The three $|\mu(\phi^0\psi^0\theta^+)>$
are degenerate with eigenvalue $\gamma(+)  = 14.383/a^3$, while
the three $|\mu(\phi^0\psi^0\theta^-)>$ are degenerate with eigenvalue
$\gamma(-) = 14.461/a^3$ (where $a=3.43$ \AA \ in AlF$_3$).
Note that $\gamma(-)$ is the largest eigenvalue of $\Gamma$, and that
$\gamma(+)$ is only $0.5\%$ smaller than $\gamma(-)$.  It follows
that the dipole-dipole interaction energy of an arbitrary 
tilt-induced dipole pattern
$|\mu(\phi^p \psi^q \theta^r)> = \sum C_{\gamma}|\gamma>$
is $-\sum |C_{\gamma}|^2 \gamma /2$.  Thus an arbitrary Glazer
tilt has dipole-dipole interaction energy
\begin{eqnarray}
E_D(\phi^p \psi^q \theta^r)&=&-\frac{1}{2}
\left(\frac{\beta e a}{2}\right)^2
\left[ \gamma(p)\phi^2+\gamma(q)\psi^2 +\gamma(r)\theta^2 \right]
\nonumber \\
&=&-\frac{1}{2}\left[\gamma(p)\mu_x^2+\gamma(q)\mu_y^2
+\gamma(r)\mu_z^2 \right]
\label{eq:glazen}
\end{eqnarray}
where $\mu_x=\beta e a \phi/2$ is the 
amplitude of the dipole eigen-array arising from
the $\hat{x}$ rotations, etc.  
When $p,q,r$ are all negative as in AlF$_3$,
the dipole-dipole interaction energy $E_D=-\gamma_D |\vec{\mu}_D|^2 /2$
is as negative as possible ($\gamma_D = \gamma(-)$ is
maximal) for any array of displacive 
dipoles of fixed magnitude $|\vec{\mu}_D|$.  
For cases when the superscripts contain some +'s, 
{\it i.e.} for different Glazer tilting schemes,
the energy is at worst 0.5\% smaller than optimum.

So far we have presented an argument showing that tilts of
corner-coupled rigid octahedra are strongly stabilized by
dipole-dipole interactions.
All such tilt systems are ``6-sublattice'' antiferroelectrics, because each
of the six anions of a BX$_6$ octahedron defines a separate direction
of polarization.  
Although it is reasonable to characterize all such dipolar-stabilized
tilt patterns as
antiferroelectric, nevertheless conventional use often \cite{Jona,Lines} 
(but not always \cite{Kanzig}) restricts 
``antiferroelectric'' to cases where an applied external field
can switch the state to ferroelectric.  Therefore we use the term
``anti-polar.''

Now we need a theory for the total energy.  From our previous
studies \cite{Chen,Chaudhuri}
it is clear that the displacive dipole $\mu_D$ must be supplemented by
an induced dipole $\mu_I$ on the polarizable anion.
This costs energy $\mu_I^2 /2\alpha$
per dipole, where $\alpha$ is the anion polarizability.
The isolated anion in a field $\vec{F}$ has a moment $\alpha \vec{F}$
which minimizes the total energy $\mu_I^2 /2\alpha-\vec{\mu}_I \cdot \vec{F}$.
For the F$^-$ anion, $\alpha$  depends somewhat on its environment,
but is in the range \cite{Kittel} $\sim 0.85 \pm$ 0.05 \AA$^3$.
We also need to account for the total change of electrostatic
energy when anions move, not just the cooperative dipole-dipole
part.  Taylor expanding the Madelung energy $\Phi$ to second order in
displacements, two types of energy appear.  
(1) There is the intersite or dipole-dipole 
term already computed which alters the energy 
bilinearly in the total moment $\mu_{I,i}
+\mu_{D,i}$ of dipoles at different sites $i$ and $j$.
(2) There is a ``Madelung'' field $\vec{\nabla}_i\Phi$ near $\vec{R}_i$
caused by the ideal undistorted lattice of other ions.  This
vanishes by inversion symmetry exactly at $\vec{R}_i$ and grows
linearly ($(\vec{u}_i \cdot \vec{\nabla}_i)\vec{\nabla}_i \Phi/2$)
with displacement from this site.  
This gives an energy
$\Gamma_{M\alpha\beta}\mu_{D\alpha}\mu_{D\beta}/2$.
Uncertainty about actual
charges is absorbed into the same factor $\beta$ which
was used in the definition
of the displacive dipole $\mu_D$.  From the Poisson equation
$\nabla^2 \Phi =0$, the trace of $\Gamma_M$ is zero.  
For cubic BX$_3$ with charges +3 for B and -1 for X, we find numerically
that $\Gamma_M$ is diagonal in cube body axes, with elements
$\gamma_M=40.789/a^3$ in the directions transverse to the B-X-B
axis (where rotations actually occur) and an element $-2\gamma_M$
in the direction parallel.  This describes a restoring field
$F_M=-\gamma_M \mu_D$ at the site of the displaced X anion nucleus.
This field is larger (by -2.8) than the destabilizing dipolar field
$F_D=\gamma_D \mu_D$ previously found from the
displacements of the other X anions.
This explains why the induced moment
($\mu_I=\alpha(F_M + F_D)$) is opposite to the displacive moments.

The energy of the dipole array generated by a tilt is
\begin{eqnarray}
U_{\rm tot}(\mu_D,\mu_I) &=& \frac{1}{2\alpha} \mu_I^2
-\frac{1}{2}\gamma_D(\mu_D + \mu_I)^2
   +\frac{1}{2}\gamma_M \mu_D^2 \nonumber \\
&+& \gamma_M \mu_D \mu_I
   -\frac{1}{2}\gamma_V \mu_D^2.
\label{eq:toten}
\end{eqnarray}
The second, third, and fourth terms contain the Taylor
expansion of the Madelung energy as described above.  The fifth
term accounts for destabilizing short-range interactions which
have not yet been discussed.  The large Madelung electrostatic
attraction of ionic crystals tries to
shrink the lattice constant as much as possible.  
Hard-core repulsion, a quantum effect, is needed 
to stabilize the lattice.  In the simplified model of 
impenetrable hard spheres, the ions touch at the 
hard-sphere radius, and the lattice constant $a$ of the
cubic phase is twice the sum of the B and X ion radii.
Pushing the X anions a distance $u$ 
off the B-X-B axis by a tilt, the lattice constant
changes to $a+\delta a$ where $\delta a = -2u^2/a$.
The Madelung energy per cell of the B$^{3+}$X$_3^{-1}$ cubic lattice is
$U_M(V)=-17.908\beta^2 e^2/V^{1/3}$ per cell, where $V=a^3$ is the
volume per cell.  When $V$ shifts to $V+\delta V$, where
$\delta V=a^2(\delta a_x +\delta a_y +\delta a_z)$, the first order shift
$U_M(V+\delta V)-U_M(V)$
under a rigid tilt by $u$ is thus $-(1/2)\gamma_V\mu_D^2$
where $\gamma_V=(4/3)17.908/a^3$ is the volume 
stabilization energy per dipole caused by the tilt $u$.

The induced
moment is the one which minimizes this energy, giving
\begin{equation}
\mu_I/\mu_D=-(\gamma_M-\gamma_D)/(1/\alpha - \gamma_D).
\label{eq:muopt}
\end{equation}
For the 
parameters of AlF$_3$, this gives $\mu_I \sim -0.80 \mu_D$.
The total energy, evaluated at this optimal choice of induced
moment, is 
\begin{equation}
U_{\rm tot}(\mu_D,\mu_{I,{\rm opt}})=\frac{1}{2}\left(\gamma_M-\gamma_D
 -\gamma_V
-\frac{(\gamma_M -\gamma_D)^2}{1/\alpha-\gamma_D} \right) \mu_D^2.
\label{eq:toten1}
\end{equation}
If we had not included the anion polarizability ({\it i.e.} 
$\alpha \rightarrow 0$) then, for the parameters of B$^{3+}$X$_3^{-1}$,
the net restoring coefficient $\gamma_M-\gamma_D -\gamma_V$ would
still be positive.  In fact, our impenetrable sphere model
probably overestimates $\gamma_V$, so stability is still fairly strong.
However, there is a critical polarizability $\alpha_c$ beyond which
the quadratic restoring energy on dipoles goes negative, given by
\begin{equation}
\frac{1}{\alpha_c}=\gamma_D + \frac{(\gamma_M-\gamma_D)^2}{\gamma_M
   -\gamma_D -\gamma_V}.
\label{eq:alphac}
\end{equation}
For our simplified model of AlF$_3$, this
is 0.136\AA$^3$.  Instability thus occurs even
for $\alpha$ well below the actual value for $F^-$, $\alpha\approx
0.85\pm 0.05$ \AA$^3$ \cite{Kittel}.

The theory presented here includes classical electrostatic
energies quite well (since non-overlapping charge distributions
interact to good approximation as if they were point charges and dipoles)
but lumps all quantum effects into a hard core, possibly
overstating the amount of energy available
in volume contraction.  
Our theory omits all higher than quadratic
effects and thus cannot predict the magnitude of the tilt.
Nevertheless, we have offered a sensible
approximation with no free parameters, which helps explain
nicely the nearly universal instability of perovskite materials
to octahedron tilting.  Furthermore, the model predicts
very little energy discrimination between different tilting schemes,
consistent with the wide range of tilts seen experimentally (and
sometimes adopted by the same material at different temperatures).
The major influence of dipole-dipole
interactions and anion polarizability are an interesting surprise.

\begin{acknowledgments}
We thank P. E. Madden, M. Wilson, and V. Perebeinos for
guidance in our earlier work which stimulated this paper. 
The work was supported in part by NSF grant no. DMR-0089492. PBA
was supported
in part by a J. S. Guggenheim Foundation fellowship.  Work at
Columbia was supported in part by the MRSEC Program of the NSF
under award no. DMR-0213574.
CPG thanks the Basic Energy Sciences program of the Department of
Energy for support via grant no. DEFG0296ER14681.
\end{acknowledgments}

\appendix*
\section{}

A generalized CM theory \cite{Allen} can be constructed as follows.
For an array of point polarizable molecules at fixed
positions $\vec{R}_i$, the energy for an arbitrary
pattern of dipoles $\{\vec{\mu}_i\}$ is
\begin{equation}
{\cal E}\left(\{\vec{\mu}_i\}\right)
=\sum_i \left(\frac{\mu_i^2}{2\alpha} 
-\frac{1}{2}\vec{\mu}_i \cdot \vec{F}_{D,i} \right)
=\frac{1}{2}<\mu|\frac{1}{\alpha} \hat{\bf 1} - {\bf{\Gamma}}|\mu>.
\label{eq:Etot} 
\end{equation}
For dipoles on the X anions in perovskite structure, there are 3 sublattice
sites per cubic unit cell.  A
vector space notation is used where $|\mu>$ is a
$9N$-dimensional column vector of the 3 Cartesian components of
each of the $3N$ dipoles in $N$ unit cells, 
and $\Gamma$ is the $9N\times 9N$
dipole-dipole interaction matrix which has elements
\begin{equation}
\Gamma_{i\alpha,j\beta}=\frac{3R_{ij\alpha}R_{ij\beta}-
\delta_{\alpha\beta}R_{ij}^2 } {R_{ij}^5}.
\label{eq:gamma}
\end{equation}
The lattice is stable against dipole formation if
the quadratic form Eq.(\ref{eq:Etot})
is positive (all its eigenvalues should be
positive.)  The condition for instability is that the maximum
eigenvalue $\gamma$ of the matrix $\Gamma$ 
exceeds the restoring force constant $1/\alpha$.
The corresponding eigenvector gives the
pattern of displacement dipoles $|\mu>$ which has
the most self-stabilizing displacment pattern.

\par
\begin{figure}
\centerline{\scalebox{0.4}[0.4]{\includegraphics{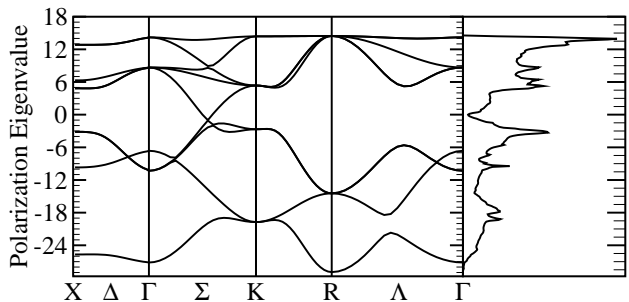}}}
\caption{\label{fig:perov}
Eigenvalues of the dipole-dipole interaction $\Gamma$ versus
wavevector for dipoles on the three F$^-$ sublattices of the
perovskite AlF$_3$ structure.  The largest eigenvalue is
the five-fold degenerate state at the R point ($\pi,\pi,\pi$),
with eigenvalue 14.461.  At the right, the density of states
is plotted horizontally versus energy vertical.}
\end{figure}
\par

Bloch's theorem allows 
eigenstates of $\Gamma$ to be chosen as simultaneous
eigenstates of translations $T(\vec{R})$
where $\vec{R}$ is any translation vector of the 
primitive (simple cubic) lattice.
The resulting Bloch states are labeled by wavevector
$\vec{k}$ which lies in the first Brillouin zone.  
The eigenvalues were computed numerically, using
an Ewald method \cite{Smith} to converge the sums. 
Results are shown in Fig. \ref{fig:perov}.
As a test of the code, eigenvalues were also calculated
for a model with a fourth sublattice corresponding to the
perovskite A sites.  Together with the three X sites,
the resulting lattice is equivalent to the face-centered
cubic structure in a non-primitive conventional cube and
a four-atom basis.  It was found numerically that the largest
eigenvalue equalled $4\pi n/3$ 
(with $n=4/a^3$) and occurred at $\vec{k}=(000)$.
This is the known result of CM theory -- when every molecule
bears the same moment $\vec{\mu}$ and sits
on a site of cubic symmetry, the field at each site is
given by the classical Lorentz value $(4\pi/3)\vec{P}$,
and a ferroelectric instability occurs when the energy
$-\vec{P}\cdot\vec{F}$ exceeds the cost $\mu^2/2\alpha$
to create the dipoles.

The flatness of the uppermost branch of $\gamma$ versus $k$ in
Fig. \ref{fig:perov} indicates that there is not much interaction
coupling the $xy$-oriented dipoles in one $xy$ plane to the
$xy$-oriented dipoles in adjacent planes.  This result
can be understood as a consequence of the exponentially
rapid transverse decay of electric field of a periodic array of dipoles
\cite{Allen}.  This fact helps explain why the observed
tilts of perovskites are so indiscriminating in their preferred
wavevector $(\pi,\pi,0)$ {\it vs.} $(\pi,\pi,\pi)$, 
and even allow mixed wavevector solutions.


\begin{thebibliography}{99}

\bibitem{Slater} J. C. Slater, Phys. Rev. {\bf 78}, 748 (1950).

\bibitem{Daniel} P. Daniel, A. Bulou, M. Rousseau, and J. Nouet,
        Phys. Rev. B {\bf 42} 10545 (1990).

\bibitem{Shannon} R. C. Shannon, Acta Cryst. {\bf A32}, 751 (1976).

\bibitem{Jackson} J. D. Jackson, {\it Classical Electrodynamics}, 3rd Ed.,
  Wiley, New York, 1999, pp 152-155.

\bibitem{Jona} F. Jona and G. Shirane, {\it Ferroelectric Crystals},
  Macmillan, New York, 1962.

\bibitem{Glazer} A. M. Glazer, Acta Cryst. B {\bf 28}, 3384 (1972).

\bibitem{Woodward} P. M. Woodward, Acta Cryst. B {\bf 53}, 32 and 44 (1997).

\bibitem{Chen} Y.-R. Chen, P. B. Allen, and V. Perebeinos, 
		Phys. Rev. B {\bf 69}, 054109 (2004).

\bibitem{Chaudhuri} S. Chaudhuri, P. Chupas, M. Wilson, P. Madden, 
		and C. P. Grey, 
		J. Phys. Chem B {\bf 108}, 3437 (2004).

\bibitem{Lines} M. E. Lines and A. M. Glass, {\it Principles and Applications
  of Ferroelectrics and Related Materials}, Clarendon Press, Oxford, 1977.

\bibitem{Kanzig} W. K\"anzig, Solid State Physics {\bf 4}, 1 (1957).

\bibitem{Kittel} C. Kittel, {\it Introduction to Solid State Physics}, 6th Ed.,
   Wiley, New York, 19xx, p.xxx.

\bibitem{Allen}	P. B. Allen, J. Chem. Phys. {\bf 120}, 2951 (2004).

\bibitem{Smith} W. Smith, CCP5 Newsletter No. 46, Oct. 1998, p.18.


\end{thebibliography}
\end{document}